 \documentclass[review,12pt]{elsarticle}

\usepackage{amssymb}

\usepackage{amsmath}
\usepackage{mathrsfs}
\usepackage{comment}
\usepackage{url}

\newcommand{\gr}[1]{\textbf{#1}}

\newcommand{\A}{\gr{A}}
\newcommand{\B}{\gr{B}}

\newcommand{\En}{\mathrm{E}}
\newcommand{\Jn}{\mathrm{J}}

\newcommand{\I}{\gr{I}}

\renewcommand{\H}{\gr{H}}
\newcommand{\E}{\gr{E}}
\newcommand{\K}{\gr{K}}

\newcommand{\W}{\gr{W}}
\newcommand{\R}{\gr{R}}
\newcommand{\Rx}{\R_{\x}}
\newcommand{\Ry}{\R_{\y}}
\newcommand{\Rn}{\R_{\n}}
\newcommand{\Rxy}{\R_{\x\y}}
\newcommand{\Ryx}{\R_{\y\x}}
\newcommand{\Rxn}{\R_{\x\n}}

\newcommand{\X}{\gr{X}}

\newcommand{\U}{\gr{U}}

\newcommand{\V}{\gr{V}}

\newcommand{\M}{\gr{M}}

\newcommand{\N}{\gr{N}}

\newcommand{\n}{\gr{n}}

\newcommand{\x}{\gr{x}}

\newcommand{\y}{\gr{y}}

\newcommand{\nn}{\mathbb{R}^{n\times n}}
\newcommand{\mm}{\mathbb{R}^{m\times m}}

\newcommand{\mn}{\mathbb{R}^{m\times n}}
\newcommand{\nm}{\mathbb{R}^{n\times m}}

\newcommand{\csn}{\mathbb{C}^N}
\newcommand{\csm}{\mathbb{C}^M}
\newcommand{\cNM}{\mathbb{C}^{N\times M}}

\newcommand{\zt}{\textrm}
\newcommand{\xr}{\mathcal{X}_r^{m\times n}}

\newcommand{\PP}{\mathcal{P}_{r}^{\iota}}

\newcommand{\ui}{\mathfrak{I}}
\newcommand{\uimal}{\iota}

\newcommand{\p}{\parallel}

\newcommand{\si}{\sigma}

\newcommand{\ep}{\epsilon}
\newcommand{\ga}{\gamma}

\newcommand{\la}{\lambda}
\newcommand{\Ga}{\boldsymbol{\Gamma}}

\newcommand{\Up}{\boldsymbol{\Upsilon}}
\newcommand{\up}{\upsilon}
\newcommand{\De}{\boldsymbol{\Delta}}

\newcommand{\de}{\delta}

\newcommand{\kd}{$\ \square$}
\newcommand{\kdm}{\ \square}

\newcommand{\no}{\noindent}
\newcommand{\pd}{\no\hspace{2em}{\itshape Proof: }}
\newcommand{\ds}{\displaystyle}

\newtheorem{theorem}{Theorem}
\newtheorem{lemma}{Lemma}

\newtheorem{remark}{Remark}
\newtheorem{fact}{Fact}

\journal{Journal of The Franklin Institute}

\begin{document}

\begin{frontmatter}



\title{Performance of the stochastic MV-PURE estimator in highly noisy settings}
\author[TP]{Tomasz Piotrowski\corref{cor1}}
\ead{tpiotrowski@is.umk.pl}
\cortext[cor1]{Corresponding author.}
\fntext[1]{Published: \url{https://doi.org/10.1016/j.jfranklin.2014.03.012}}
\address[TP]{Department of Informatics,\\
  Faculty of Physics, Astronomy and Informatics,\\
  Nicolaus Copernicus University,\\
  Grudziadzka 5, 87-100 Torun, Poland}

\author[IY]{Isao Yamada}
\ead{isao@sp.ce.titech.ac.jp}
\address[IY]{Department of Communications and Computer Engineering,\\
  Tokyo Institute of Technology,\\
  Tokyo 152-8550, Japan}


\begin{abstract}
The stochastic minimum-variance pseudo-unbiased reduced-rank estimator (stochastic MV-PURE estimator) has been developed to provide linear estimation with robustness against high noise levels, imperfections in model knowledge, and ill-conditioned systems. In this paper, we investigate the theoretical performance of the stochastic MV-PURE estimator under varying levels of additive noise. We prove that the mean-square-error (MSE) of this estimator in the low signal-to-noise (SNR) region is much smaller than that obtained with its full-rank version, the minimum-variance distortionless estimator, and the gap becomes larger as the noise level increases. These results shed light on the excellent performance of the stochastic MV-PURE estimator in highly noisy settings obtained in simulations so far. Furthermore, we extend previous numerical simulations to show how the insight gained from the results of this paper can be used in practice.
\end{abstract}


\end{frontmatter}


\section{Introduction}
\label{intro}
Linear estimation of an unknown random vector of parameters in linear regression models has been a subject of continuous research, amplified by the widespread use of the linear model $\y=\H\x+\sqrt{\ep}\n$, e.g., in brain signal processing \cite{Cichocki2002,Piotrowski2013}, wireless communications \cite{Wang2004}, and array signal processing \cite{VanTrees2002}, \cite[Sec.1.2]{Pezeshki2010}. In particular, the search for linear estimators with greater robustness against high noise levels, imperfections in model knowledge, and ill-conditioned systems than the theoretically optimal [in the mean-square-error     (MSE) sense] linear minimum mean-square-error (MMSE) estimator (Wiener filter) \cite{Luenberger1969,Kailath2000} has long attracted researchers' and practitioners' attention, see e.g., \cite{Huber1964,Kassam1985,Eldar2004,Eldar2005,Rong2005} for an overview and recent solutions. A natural way to increase the robustness of an estimator is to formulate it as a solution of a MSE minimization problem with carefully chosen additional constraints. Such estimator is theoretically suboptimal when compared with the MMSE estimator, but it achieves better performance in certain settings of practical interest (e.g., if only a finite sample estimate $\widehat{\H}$ of a model matrix $\H$ is available in the linear model $\y=\H\x+\sqrt{\ep}\n$). An example of this approach is an estimator $\W$ minimizing MSE subject to the so-called
distortionless constraint $\W\H=\I$ \cite{Wang2004,Shahbazpanahi2004}. This estimator is essentially equivalent to the celebrated best linear unbiased estimator (BLUE estimator)~\cite{Kailath2000} in deterministic estimation.\footnote{Due to this equivalence, we will call this   estimator as stochastic BLUE throughout this paper.}

Another provably robust technique is the reduced-rank approach \cite{Brillinger1975,Scharf1991,Stoica1996,Scharf1998,deLamare2012,Huang2012}, very useful in ill-conditioned and highly noisy settings. Thus, it would seem natural to extend the distortionless approach of the stochastic BLUE estimator to the reduced-rank case in order to achieve further robustness against high noise levels, imperfections in model knowledge, and ill-conditioned systems. This has been achieved by the  introduction of the \emph{minimum-variance pseudo-unbiased reduced-rank estimator} (MV-PURE estimator) proposed for the deterministic case in \cite{Yamada2006, Piotrowski2008}, and extended to the stochastic case in~\cite{Piotrowski2009}. Focusing on the stochastic case, the stochastic MV-PURE estimator is derived as a solution of the following hierarchical nonconvex constrained optimization problem. In the first stage optimization, under a rank constraint, we minimize simultaneously all unitarily invariant norms of $\W\H-\I$ with the objective of suppressing the distortion caused by the (reduced-rank) estimator. Then, in the second stage optimization, among all solutions of the first stage optimization, we find the one achieving minimum~MSE.

To compare the robustness properties of the estimators considered in this paper, we first compare their performances in the theoretical case of perfect model knowledge. Then, aided with theoretical results serving as a benchmark, one would be able to determine the performance degradation of a given estimator (i.e., its robustness properties) in the presence of model uncertainties. To this end, this study introduces conditions under which the rank reduction technique of the stochastic MV-PURE estimator enables it (in the theoretical setting of perfect model knowledge) to achieve significantly lower MSE than its full-rank version, the stochastic BLUE estimator. The main result of this paper shows that the larger the power of the additive noise, the larger the gap in performance between the stochastic MV-PURE and stochastic BLUE estimators. In addition, we also show that the optimal rank is in fact monotonically decreasing with increasing noise level. These findings confirm the common intuition that insistence on distortionless is an inadequate requirement in highly noisy conditions. The results of this paper derive conditions when this intuition is true. Furthermore, they demonstrate clearly the interplay between noise level and ill-conditioning of the model considered. 

A numerical example is provided to demonstrate insight provided by the results of this paper in practical applications. We extend simulations considered first in \cite{Piotrowski2009}, where the stochastic MV-PURE estimator was employed as a linear receiver for multiple-input multiple-output (MIMO) wireless communication system.

Preliminary short versions of this paper have been presented at conferences \cite{Piotrowski2008b,Piotrowski2009b}.

\section{Preliminaries}
\subsection{Stochastic linear model} \label{slm}
In this paper the following stochastic linear model is considered \cite{Luenberger1969,Kailath2000}:
\begin{equation} \label{general}
  \y=\H\x+\sqrt{\ep}\n,
\end{equation}
where $\y, \x, \n$ represent an observable random vector, a random vector to be estimated, and additive noise, respectively. Furthermore, $\H\in\nm$ is a known matrix of rank $m$, and $\ep>0$ is a known constant. It is assumed that $\x$ and $\n$ have zero mean and are uncorrelated, $\Rxn=0$, and that $\Rx\succ 0$ and $\Rn\succ 0$ are known positive definite covariance matrices of $\x$ and $\n$, respectively (by $\A\succ 0$ we mean that matrix $\A$ is positive definite). From the previous assumption, $\Ry=(\H\Rx \H^t+\ep\Rn)\succ 0$ and $\Rxy=\Rx \H^t$ are available and $rank(\Rxy)=m$. The norm of a random vector $\x$ is defined by $\p\x\p=\sqrt{tr\left[\En(\x\x^t)\right]}$, and, without loss of generality, it is assumed that $tr[\Rn]=1$ so that $\p\sqrt{\ep}\n\p^2=\ep.$

We consider the problem of linear estimation of $\x$ given $\y$ with the mean-square-error (MSE) as the performance criterion. Thus, one seeks a fixed matrix $\W\in\mn$, called here an estimator, for which an estimate of $\x$ given by 
\begin{equation}
  \widehat{\x}=\W\y
\end{equation}
is optimal with respect to a certain criterion related to the MSE of~$\widehat{\x}$:
\begin{equation} \label{mse}
  \Jn(\W)=\p\widehat{\x}-\x\p^2=tr[\W\Ry \W^t]-2tr[\W\Ryx]+tr[\Rx].
\end{equation}

\subsection{MMSE, stochastic BLUE, and stochastic MV-PURE estimators}
The unique solution of the problem of minimizing (\ref{mse}) is the linear minimum mean-square-error estimator (MMSE) $\W_{MMSE}$, often called the Wiener filter, given by \cite{Luenberger1969,Kailath2000}
\begin{equation} \label{mmse}
  \W_{MMSE}=\Rxy\Ry^{-1}=\Rx \H^t(\H\Rx \H^t+\ep\Rn)^{-1}.
\end{equation}

As discussed in section \ref{intro}, the search for linear estimators under the MSE criterion does not end here. Certain solutions are more robust against high noise levels, imperfections in model knowledge, and ill-conditioned systems. A popular approach is to introduce the so-called distortionless constraint $\W\H=\I_m$ to the MSE minimization problem \cite{Kailath2000}. The solution of this problem is the stochastic BLUE estimator which is often called the distortionless MMSE estimator and has been widely employed in recent signal estimation and detection schemes \cite{Wang2004,Shahbazpanahi2004}. Namely, the stochastic BLUE estimator is defined as the unique solution of the following optimization problem:
\begin{equation} \label{blue}
  \left\{
  \begin{array}{ll}
    \zt{minimize} & \Jn(\W)\vspace{0.1cm}\\
    \zt{subject to} & \W\H=\I_m,\\
  \end{array}\right.
\end{equation}
with the unique solution\footnote{Note that in the deterministic case, the condition $\W\H=\I_m$ in   (\ref{blue}) implies unbiasedness of the estimator $\W_{BLUE}$   (\ref{blueish}).} 
\begin{equation} \label{blueish}
  \W_{BLUE}=(\H^t\Ry^{-1}\H)^{-1}\H^t\Ry^{-1}=(\H^t\Rn^{-1}\H)^{-1}\H^t\Rn^{-1}.
\end{equation}

The stochastic MV-PURE estimator introduced in \cite{Piotrowski2009} is a reduced-rank extension of the approach (\ref{blue}), as it achieves the minimum distortion among all reduced-rank estimators. More precisely, the stochastic MV-PURE estimator is defined as the solution of the following optimization problem for a given rank constraint~$r\leq m$:
\begin{equation} \label{mvpulreS}
  \left\{
  \begin{array}{ll}
    \zt{minimize} & \Jn(\W_r)\\
    \zt{subject to} & \ds \W_r\in\bigcap_{\uimal\in\ui}\PP,\\
  \end{array}\right.
\end{equation}
where
\begin{equation} \label{argminS}
  \PP=\arg\min_{\W_r\in\xr}\p \W_r\H-\I_m\p_\uimal^2,\ \uimal\in\ui,
\end{equation}
$\xr:=\{\W_r\in\mn: rank(\W_r)\leq r\leq m\}$, and $\ui$ is the index set of all unitarily invariant norms. Let us recall here that a matrix norm~$\uimal$ is unitarily invariant if it satisfies $\p \U\X\V\p_\uimal=\p \X\p_\uimal$ for all orthogonal $\U\in\mm,\ \V\in\nn$ and all $\X\in\mn$ \cite{Horn1985} [the Frobenius, spectral, and trace (nuclear) norms are examples of unitarily invariant norms]. The following theorem provides a closed algebraic form of the stochastic MV-PURE estimator.

\begin{theorem}[$\mbox{\cite{Piotrowski2009}}$] \label{glt2}
Consider the optimization problem (\ref{mvpulreS}). The following     holds:
  \begin{enumerate}
  \item Assume that the rank $r$ is constrained to $r<m$, and define         the symmetric matrix $\K\in\mm$ by
    \begin{equation} \label{K}
      \K:=\left(\H^t\Ry^{-1}\H\right)^{-1}-2\Rx.
    \end{equation}
Let the eigenvalue decomposition of the symmetric matrix $\K$ be given by $EVD(\K)=\E\De \E^t$, with eigenvalues       $\de_1\leq\de_2\leq\dots\leq\de_{m}$ organized in nondecreasing order,         and $e_j$ denotes an eigenvector associated with the eigenvalue         $\de_j$. Then, the solution of the problem (\ref{mvpulreS}), denoted $\W^r_{sMV-PURE}\in\mn$, is given by
    \begin{equation} \label{mvpform}
      \W^r_{sMV-PURE}=\E_{r}\E_{r}^t\W_{BLUE},
    \end{equation}
    where $\W_{BLUE}$ is defined in (\ref{blueish}) and         $\E_{r}:=(e_{1},\ldots,e_r).$ If $\de_r\neq\de_{r+1}$, the solution is         unique. Moreover:
    \begin{equation} \label{mvpvalue}
      \Jn(\W^r_{sMV-PURE})=\sum_{i=1}^r\de_i+tr[\Rx].
    \end{equation}
  \item If no rank constraint is imposed, i.e., if $r=m$, then the         solution of the problem (\ref{mvpulreS}) is uniquely given by         $\W^m_{sMV-PURE}=\W_{BLUE}$. In particular:
    \begin{equation} \label{mvpvalue2}
      \Jn(\W^m_{sMV-PURE})=\Jn(\W_{BLUE})=\sum_{i=1}^m\de_i+tr[\Rx].
    \end{equation}
  \end{enumerate}
\end{theorem}
\vspace{.3cm}

At this stage, it is useful to remark on optimization problems (\ref{blue}) and (\ref{mvpulreS}) (for $r<m$), and their respective solutions (\ref{blueish}) and (\ref{mvpform}). It is simple to verify that the estimates of~$\x$ obtained in model (\ref{general}) by the stochastic BLUE estimator and the stochastic MV-PURE estimator of rank $r<m$ are
\begin{equation} \label{blue_out}
\W_{BLUE}\y=\W_{BLUE}\H\x+\W_{BLUE}\sqrt{\ep}\n=\x+\W_{BLUE}\sqrt{\ep}\n,
\end{equation} 
and
\begin{multline} \label{mvp_out}
\W^r_{sMV-PURE}\y=\W^r_{sMV-PURE}\H\x+\W^r_{sMV-PURE}\sqrt{\ep}\n=
\\\E_{r}\E_{r}^t\x+\E_{r}\E_{r}^t\W_{BLUE}\sqrt{\ep}\n,
\end{multline} 
respectively. Then, the minimum distortion property of the stochastic MV-PURE estimator is seen by observing that in virtue of the constraints employed in optimization problem (\ref{mvpulreS}) one has
\begin{equation} \label{mindis}
\p \W^r_{sMV-PURE}\H-\I_m\p_\uimal=\p\E_{r}\E_{r}^t-\I_m\p_\uimal,
\end{equation}
which minimizes the distance between $\I_m$ and any matrix of rank at most $r<m$ for any unitarily invariant norm $\uimal$, in view of the Mirsky-Schmidt Approximation Theorem \cite{Mirsky1960}. Therefore, in the sense of the set of equalities (\ref{mindis}), the stochastic MV-PURE estimator induces in (\ref{mvp_out}) a minimum distortion on the reconstructed vector $\x$ by orthogonally projecting it onto the subspace spanned by columns of $\E_{r}$, for a given rank constraint $r<m.$ Indeed, the exact conditions for (\ref{mvp_out}) to achieve lower MSE than (\ref{blue_out}) if the power $\ep$ of the additive noise is sufficiently large will be given in the following section. 

We also note the following remark, which follows directly from theorem~\ref{glt2}.

\begin{remark} \label{remake}
  Let us set the rank constraint $r<m.$ Then:
  \begin{eqnarray} \label{gen}
    \Jn(\W_{sMV-PURE}^r)<\Jn(\W_{sMV-PURE}^{r+1})<\nonumber\\\dots<\Jn(\W_{sMV-PURE}^m)=\Jn(\W_{BLUE})
    \Longleftrightarrow\de_{r+1}>0.
  \end{eqnarray}
  In particular, the optimal rank $r_{opt}$ of the stochastic MV-PURE           estimator, for which $\Jn(\W_{sMV-PURE}^{r_{opt}})$ is the smallest, is such that
  \begin{equation} \label{gek}
    r_{opt}=\max\left\{s\in \{1,2,\ldots,m\} \mid         \delta_{s}<0\right\}.     \kdm
  \end{equation}
\end{remark}

\subsection{Variational characterization of eigenvalues of symmetric matrices} \label{kru}
The following theorem gives two-sided bounds for the eigenvalues of $A+B$ for any symmetric matrices $A$ and $B.$ This result will play an important role in derivation of the results of section~\ref{main}.
\begin{fact}[$\mbox{Weyl \cite[p.181]{Horn1985}}$] \label{wenflon}
  Let $A,B\in\mm$ be symmetric matrices, and let the eigenvalues     $\la_i(A)$, $\la_i(B)$, and $\la_i(A+B)$ be organized in nondecreasing     order. For each $k=1,2,\dots,m$ we have:
  \begin{equation} \label{welon}
    \la_k(A)+\la_1(B)\leq\la_k(A+B)\leq\la_k(A)+\la_m(B).
  \end{equation}
\end{fact}

\section{Performance analysis under varying levels of signal-to-noise\\ ratio} \label{main}
The following lemma gives an alternative expression of the matrix K defined in~(\ref{K}). 
\begin{lemma} \label{tolpa}
With notation as in model (\ref{general}), the following equality     holds:
  \begin{equation} \label{eqieqi}
    (\H^t\Ry^{-1}\H)^{-1}=\ep(\H^t\Rn^{-1}\H)^{-1}+\Rx,
  \end{equation}
  and thus $\K$ (\ref{K}) in theorem \ref{glt2} can be expressed as:
  \begin{equation} \label{Kalt}
    \K=\left(\H^t\Ry^{-1}\H\right)^{-1}-2\Rx=\ep\left(\H^t\Rn^{-1}\H\right)^{-1}-\Rx.
  \end{equation}
\end{lemma}
\pd Let us recall that in model (\ref{general}) we have $\Rxy=\Rx \H^t$ (and therefore $\Ryx=\H\Rx$), and consider the covariance matrix
\begin{equation} \label{cov}
  \E\left[(\W\y-\x)(\W\y-\x)^t\right]=\W\Ry \W^t-\W\H\Rx-\Rx \H^t\W^t+\Rx.
\end{equation}
By inserting the first expression of $\W_{BLUE}=(\H^t\Ry^{-1}\H)^{-1}\H^t\Ry^{-1}$ (\ref{blueish}) into~(\ref{cov}), we obtain $\E\left[(\W_{BLUE}\y-\x)(\W_{BLUE}\y-\x)^t\right]=(\H^t\Ry^{-1}\H)^{-1}-\Rx.$ However, since $\Ry=\H\Rx \H^t+\ep\Rn$, from the second expression of $\W_{BLUE}=(\H^t\Rn^{-1}\H)^{-1}\H^t\Rn^{-1}$ (\ref{blueish}), we obtain from (\ref{cov}) also the alternative expression $\E\left[(\W_{BLUE}\y-\x)(\W_{BLUE}\y-\x)^t\right]=\ep(\H^t\Rn^{-1}\H)^{-1}.$ \kd
\vspace{.3cm}

The following theorem demonstrates the usefulness of the reduced-rank approach of the stochastic MV-PURE estimator in highly noisy settings.

\begin{theorem} \label{battery}
Let $EVD(\H^t\Rn^{-1}\H)=\M\Up \M^t$ be the eigenvalue decomposition of 
\begin{equation}
\H^t\Rn^{-1}\H\succ 0,
\end{equation}
with eigenvalues $\up_1\geq\up_2\geq\dots\geq\up_m>0.$ Moreover, let $EVD(\Rx)=\N\Ga\N^t$ be the eigenvalue decomposition of $\Rx\succ 0$ with eigenvalues $\ga_1\geq\ga_2\geq\dots\geq\ga_m>0.$ Then, for each rank constraint $r<m$, if the power $\ep$ of the additive noise is such that:
  \begin{equation} \label{ulala}
    \ep>\up_{r+1}\ga_1,
  \end{equation}
  then:
  \begin{eqnarray} \label{elala}
    \Jn(\W_{sMV-PURE}^r)<\Jn(\W_{sMV-PURE}^{r+1})<\nonumber\\
\dots<\Jn(\W_{sMV-PURE}^m)=\Jn(\W_{BLUE}).
  \end{eqnarray}
In particular, (\ref{ulala}) and (\ref{elala}) guarantee that the reduced-rank approach of the stochastic MV-PURE estimator enables it to achieve lower mean-square-error than the stochastic BLUE estimator for all~$\ep>\up_m\ga_1.$ Moreover, if:
  \begin{equation} \label{alala}
    \up_r\ga_m>\ep>\up_{r+1}\ga_1,
  \end{equation}
  then $r=r_{opt}=\max\left\{s\in \{1,2,\ldots,m\} \mid     \delta_{s}<0\right\}.$
\end{theorem}
\pd From our assumptions, we have $EVD[(\H^t\Rn^{-1}\H)^{-1}]=\M\Up^{-1}\M^t$ with eigenvalues $0<\up_1^{-1}\leq\up_2^{-1}\leq\dots\leq\up_m^{-1}$, and $EVD(-\Rx)=\N(-\Ga)\N^t$ with eigenvalues $-\ga_1\leq-\ga_2\leq\dots\leq-\ga_m<0.$
Thus, if we denote the eigenvalue decomposition of $\K=\ep\left(\H^t\Rn^{-1}\H\right)^{-1}-\Rx$ (\ref{Kalt}) by $EVD(\K)=\E\De \E^t$, with eigenvalues $\de_1\leq\de_2\leq\dots\leq\de_{m}$ organized in nondecreasing order, from fact~\ref{wenflon} in section \ref{kru}, upon setting
$\A=\ep(\H^t\Rn^{-1}\H)^{-1}$ and $\B=-\Rx$, we obtain for each $s=1,2,\dots,m$:
\begin{equation} \label{low}
  \ep\up_s^{-1}-\ga_1\leq\de_s\leq\ep\up_s^{-1}-\ga_m.
\end{equation}
Therefore, from the first inequality above it is seen that condition (\ref{ulala}) ensures $\de_{r+1}>0$, which in view of (\ref{gen}) is equivalent to (\ref{elala}). Similarly, if $\ep$ satisfies the more stringent condition (\ref{alala}), then not only $\de_{r+1}>0$, but also $\de_r<0$, which in view of (\ref{gek}) implies that~$\ds r=r_{opt}.$ \kd

It should be noted that from the second inequality in (\ref{low}) one has for $s=m$ that $\de_m\leq\ep\up_m^{-1}-\ga_m.$ This implies in particular that if $\ep<\up_m\ga_m$, then $\de_m<0$, which in view of (\ref{gek}) implies that $m=r_{opt}.$ This fact justifies our focus on the `highly noisy' settings in this paper, in the well-defined sense introduced by theorem \ref{battery}. 

Theorem \ref{battery} gives also an insight into the interplay between the noise power $\ep$ and ill-conditioning of $\H^t\Rn^{-1}\H.$ Namely, for fixed $\Rx\succ 0$, if $\H^t\Rn^{-1}\H$ possesses some vanishingly small trailing eigenvalues $\up_{r+1},\dots,\up_m$ for some $r<m$, then in such settings it suffices for the noise power $\ep$ to be relatively small for the reduced-rank approach to be useful, in the sense of the relation given by (\ref{ulala})-(\ref{elala}).

Moreover, theorem \ref{battery} can be generalized by exchanging the roles of $\A$ and~$\B$ in the proof of theorem \ref{battery}, which would give (in virtue of fact \ref{wenflon} in section \ref{kru}) alternative two-sided bounds for the eigenvalues of $\K.$ Indeed, many different two-sided bounds for the eigenvalues of $K$ can be obtained by applying the more general theorem 4.3.7 in \cite[pp.184-185]{Horn1985}, fact \ref{wenflon} in section \ref{kru} being a special case. Nevertheless, the two-sided bounds in (\ref{low}), used in the proof of theorem~\ref{battery}, give the naturally interpretable conditions (\ref{ulala}) and (\ref{alala}).

However, the conditions in (\ref{alala}) do not guarantee that the optimal rank of the stochastic MV-PURE estimator decreases as the noise power $\ep$ increases when the input vector $\x$ is not necessarily white. This fact is obtained in the following theorem.

\begin{theorem} \label{blackened}
For fixed $\Rx\succ 0$ and $\Rn\succ 0$ in the stochastic linear model in Sec.\ref{slm}, the optimal rank of the stochastic MV-PURE estimator (for which it achieves the smallest MSE among all rank constraints) is monotonically decreasing with increasing power $\ep$ of the additive noise.
\end{theorem}
\pd Denote the eigenvalue decomposition of $\K=\ep\left(\H^t\Rn^{-1}\H\right)^{-1}-\Rx$ (\ref{Kalt}) by $EVD(\K)=\E\De \E^t$, with eigenvalues $\de_1\leq\de_2\leq\dots\leq\de_{m}$ organized in nondecreasing order. We recall from remark~\ref{remake} that the rank constraint $r_{opt}$ is optimal in our sense if it satisfies $r_{opt}=\max\left\{s\in \{1,2,\ldots,m\} \mid \delta_{s}<0\right\}.$ Now we show that all eigenvalues of $\K$ grow monotonically with increasing $\ep$, which is a sufficient condition for the claim in the theorem, because in such a case $r_{opt}$ can only be monotonically decreasing with increasing~$\ep.$

To this end, denote by $\la_k(\X)$ the eigenvalues of a given symmetric matrix~$\X$ organized in nondecreasing order and consider $\ep_1>\ep_0>0$ so that 
\begin{equation}
\ep_1(\H^t\Rn^{-1}\H)^{-1}-\Rx=\A+\B,
\end{equation}
where $\A=\ep_0(\H^t\Rn^{-1}\H)^{-1}-\Rx$ and $\B=\ep_2(\H^t\Rn^{-1}\H)^{-1}$ with $\ep_2=\ep_1-\ep_0>0.$ Note that $\B$ so defined is a positive definite matrix, thus all eigenvalues of $\B$ are strictly positive. Hence, from the first inequality in fact \ref{wenflon} in section~\ref{kru}, we obtain that
\begin{eqnarray}
  \forall     k=1,\dots,m\quad\la_k\left[\ep_1(\H^t\Rn^{-1}\H)^{-1}-\Rx\right]\geq
  \la_k\left[\ep_0(\H^t\Rn^{-1}\H)^{-1}-\Rx\right]+\nonumber\\\la_1\left[\ep_2(\H^t\Rn^{-1}\H)^{-1}\right]
  >\la_k\left[\ep_0(\H^t\Rn^{-1}\H)^{-1}-\Rx\right],\quad
\end{eqnarray}
which completes the proof. \kd

\section{Numerical example} \label{ne}
In digital signal processing applications, the following setup is frequently considered:
\begin{equation} \label{mul}
\y_c=\H_c\x_c+\sqrt{\ep}\n_c,
\end{equation}
where $\H_c\in\cNM$ is a complex Gaussian channel between input signal $\x_c\in\csm$ and output signal $\y_c\in\csn$ corrupted by additive Gaussian noise $\n_c\in\csn.$ The task is to estimate $\x_c$ by $\widehat{\x}_c=\W\y_c$ under the MSE criterion. 

We set $N=M=8$ and assume zero-mean temporally white circular Gaussian noise random vector with spatial covariance matrix $\R_{\n_c}\succ 0.$ The entries of $\H_c$ are i.i.d. drawn from a Gaussian distribution with zero-mean and unity variance. The input signal $\x_c$ consists of symbols drawn uniformly and independently from the QPSK=$\{1+i,1-i,-1+i,-1-i\}$ constellation, and it is uncorrelated with noise, thus $\R{\x_c\n_c}=0\in\mathbb{R}^{M\times N}.$ The signal-to-noise ratio ($SNR$) is defined as: 
\begin{equation} \label{SNR}
\textrm{SNR}={{\si^2_h}\over{\ep tr[R_{\n_c}]}}={{1}\over{\ep}},
\end{equation}
where $\si^2_h=1$ is the variance of the elements of $\H_c$, and where, without loss of generality, it is assumed that $tr[\R_{\n_c}]=1.$ The noise covariance matrix $\R_{\n_c}$ is the same throughout the remainder of section \ref{ne}, and it is assumed below that $Q=200$ data blocks have been transmitted.

The effectiveness of the stochastic MV-PURE estimator was demonstrated in \cite{Piotrowski2009}, where it was employed as a linear receiver for multiple-input multiple-output (MIMO) wireless communication system. However, it was unclear how the power of the additive noise and the ill-conditioning of the system considered affected the performance of the stochastic MV-PURE estimator, especially compared to its full-rank version, the stochastic BLUE estimator. Thanks to the results of section \ref{main}, these questions can be answered below. 

To this end, let us first cast the complex model (\ref{mul}) into its equivalent real-valued representation:
\begin{equation} \label{under}
\y=\left(
\begin{array}{c} 
Re({\y_c})\\
Im({\y_c})\\
\end{array}\right)\!\!,\ 
\n=\left(
\begin{array}{c} 
Re({\n_c})\\
Im({\n_c})\\
\end{array}\right)\!\!,\ 
\x=\left(
\begin{array}{c} 
Re({\x_c})\\
Im({\x_c})\\
\end{array}\right)\!\!,
\end{equation}
and:
\begin{equation} \label{line}
\H=\left(
\begin{array}{cc} 
Re({\H_c}) & -Im({\H_c})\\
Im({\H_c}) & Re({\H_c})\\
\end{array}\right)\!\!,\ 
\Rn=1/2\left(
\begin{array}{cc} 
R_{\n_c} & 0\\
0 & R_{\n_c}\\
\end{array}\right)\!\!,
\end{equation}
with $\Rx=\I_{16}.$ The sample mean estimate of the mean-square-error~(\ref{mse}) is given by:
\begin{equation} \label{wJ}
\widehat{\Jn(\W)}={{1}\over{Q}}\sum_{q=1}^Q\p\widehat{\x}_{c_q}-\x_{c_q}\p^2,
\end{equation}
where $Q$ is the number of data blocks, and $\widehat{\x}_{c_q}$ is the estimate of $\x_{c_q}$ obtained from its real-valued representation $\widehat{\x}_q=\W\y_q$ as $\widehat{\x}_{c_q}=[\I_8\quad  i\I_8]\widehat{\x}_q.$ In Fig.\ref{fig:1} and Fig.\ref{fig:3} below, the sample mean estimate of the mean-square-error is represented in decibels [$10\log_{10}\widehat{\Jn(\W)}$] as $MSE[dB].$ The following levels of $SNR[dB]$ are considered below:
\begin{equation}
SNR[dB]=(-8,-6,-4,-2,0,2,4,6,8),
\end{equation}
which correspond to $\ep$ values of 
\begin{equation} \label{eps}
\ep=(6.31,3.98,2.51,1.58,1,0.63,0.4,0.25,0.16),
\end{equation}
respectively, \emph{cf.} (\ref{SNR}).

\begin{figure}[ht]
  \includegraphics[scale=.506]{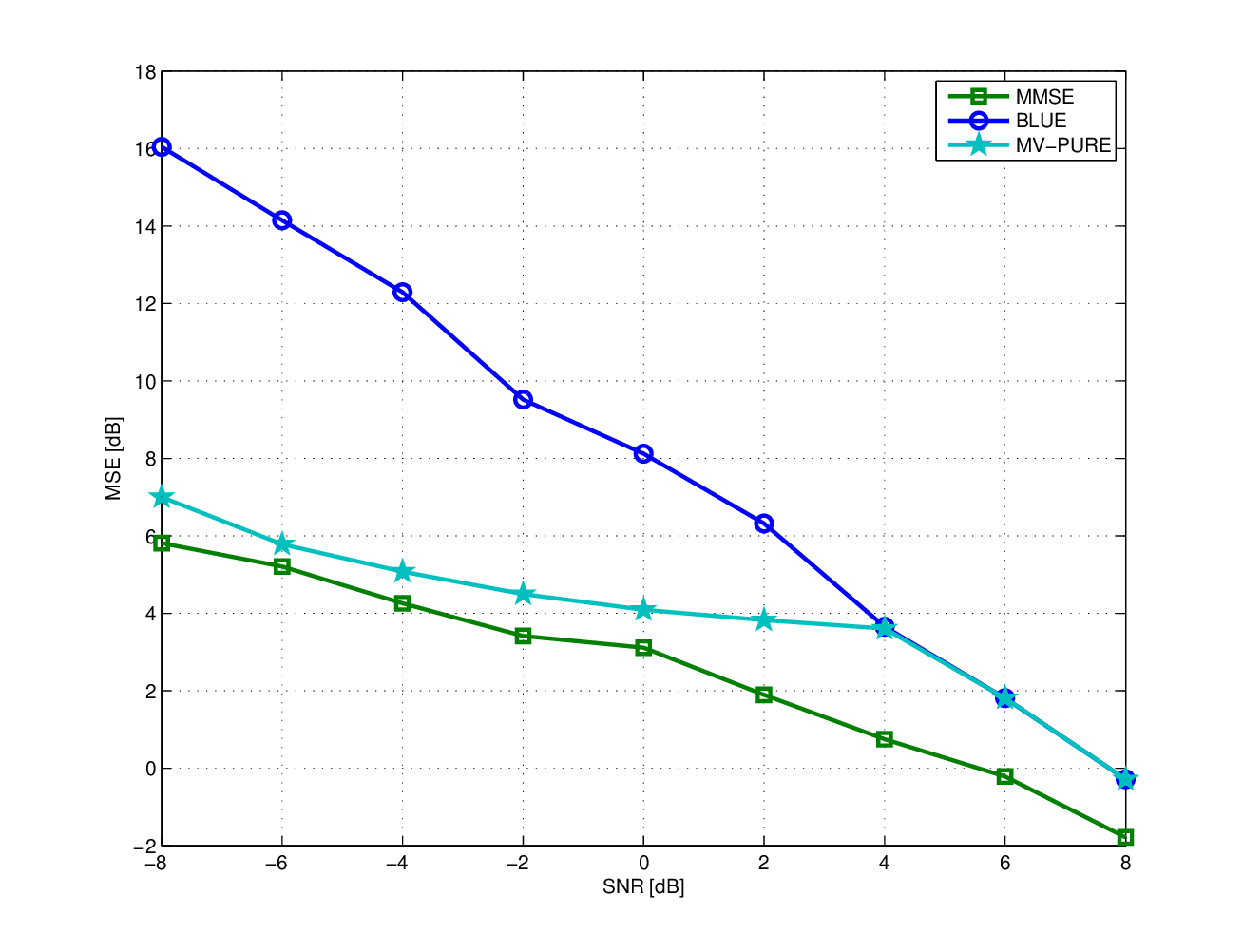}
\caption{MSE [dB] vs. SNR [dB] for a sample channel realization in theoretical case.}
\label{fig:1}
\end{figure}

In the theoretical case of exact model knowledge, a performance comparison for a sample realization of channel $\H_c$ is presented in Fig.\ref{fig:1}, where the eigenvalues of $\H^t\Rn^{-1}\H$ are found to be:
\begin{multline} \label{eig1} (\up_1=9427,\up_2=9427,\up_3=1716.8,\up_4=1716.8,\dots,\\\up_{13}=6.55,\up_{14}=6.55,\up_{15}=0.35,\up_{16}=0.35).
\end{multline}
Note that the eigenvalues of $\H^t\Rn^{-1}\H$ come in pairs in virtue of the real-valued representation via (\ref{under})-(\ref{line}). Therefore, using theorem \ref{battery} for $\Rx=\I_{16}$ (thus $\ga_i=1$ for $i=1,\dots,16$ in theorem \ref{battery}) and for the channel realization as in Fig.\ref{fig:1}, from (\ref{eps}) and (\ref{eig1}) we obtain that $r_{opt}=14$ for $SNR[dB]=(-8,-6,-4,-2,0,2,4)$, and in view of the discussion below theorem \ref{battery} we have that $r_{opt}=16$ for $SNR[dB]=(6,8).$ 

Moreover, from theorem \ref{glt2} and remark \ref{remake}, it is simple to verify that for $\Rx=\I_{16}$ one has from (\ref{gek}) that 
\begin{equation} \label{r15}
r_{opt}=\max\left\{s\in \{1,2,\ldots,m\} \mid \si_s>0.5\right\},
\end{equation}
where $\si_1\geq\si_2\geq\dots\geq\si_{16}>0$ are the eigenvalues of $\H^t\Ry^{-1}\H.$ Thus, theorem \ref{blackened} implies that, if $\si_{14}>0.5$ and $\si_{15}<0.5$ for certain $SNR[dB]$ levels $x,y\in\{-8,-6,-4,-2,0,2,4,6,8\}$ with $y<x$, then one can set $r_{opt}=14$ without any numerical simulations for all levels of $SNR[dB]$ such that $$y\leq SNR[dB]\leq x.$$ In particular, $\si_{15}$ must be monotonically decreasing with decreasing levels of $SNR[dB]$, as demonstrated in Fig.\ref{fig:2} below. This can be deducted from the proof of theorem \ref{blackened}, which shows that all eigenvalues of $\K=\left(\H^t\Ry^{-1}\H\right)^{-1}-2\Rx$ (which may be expressed here as  $\si_1^{-1}-2\leq\si_2^{-1}-2\leq\dots\leq\si_{16}^{-1}-2$) grow monotonically with decreasing levels of $SNR[dB].$ 

\begin{figure}[ht]
  \includegraphics[scale=.6]{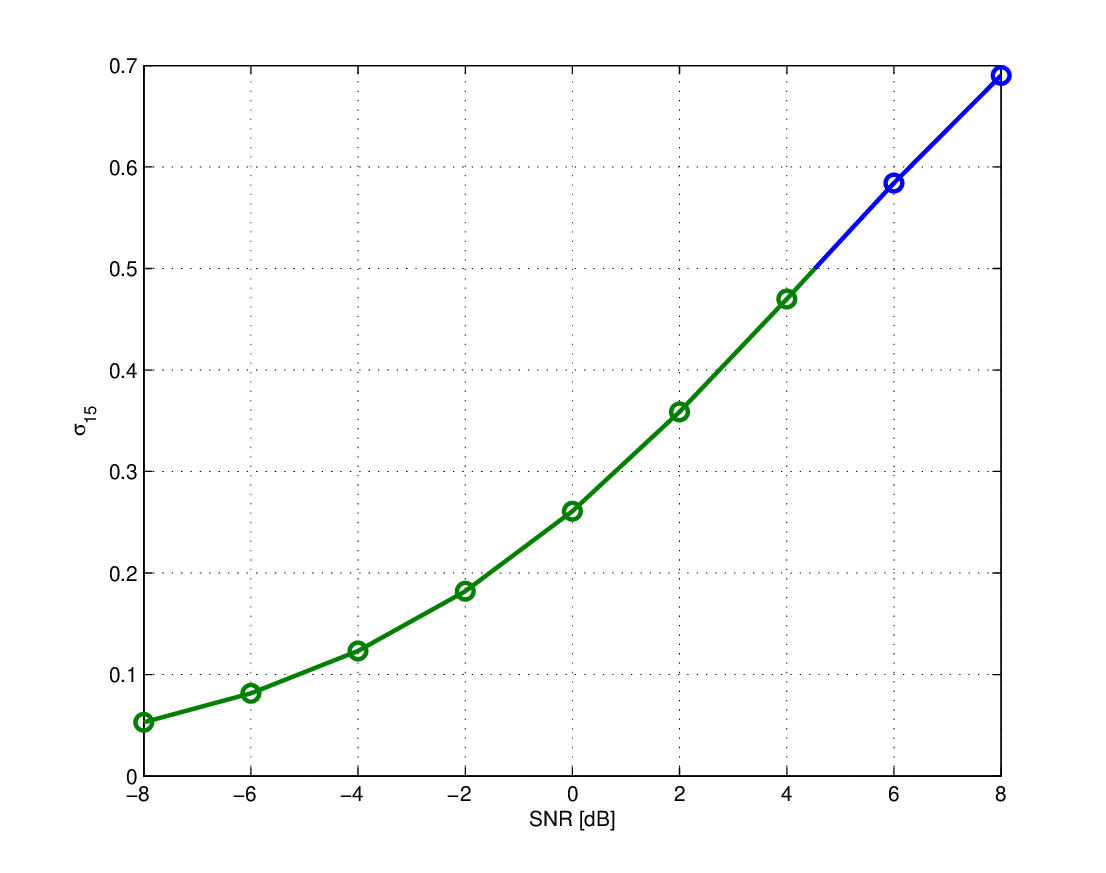}
\caption{Monotonic decrease of $\si_{15}$ with decreasing $SNR[dB]$ in theoretical case. The 0.5 threshold is crossed between $SNR[dB]=4$ and $SNR[dB]=6.$}
\label{fig:2}
\end{figure}

\begin{figure}[ht]
  \includegraphics[scale=.6]{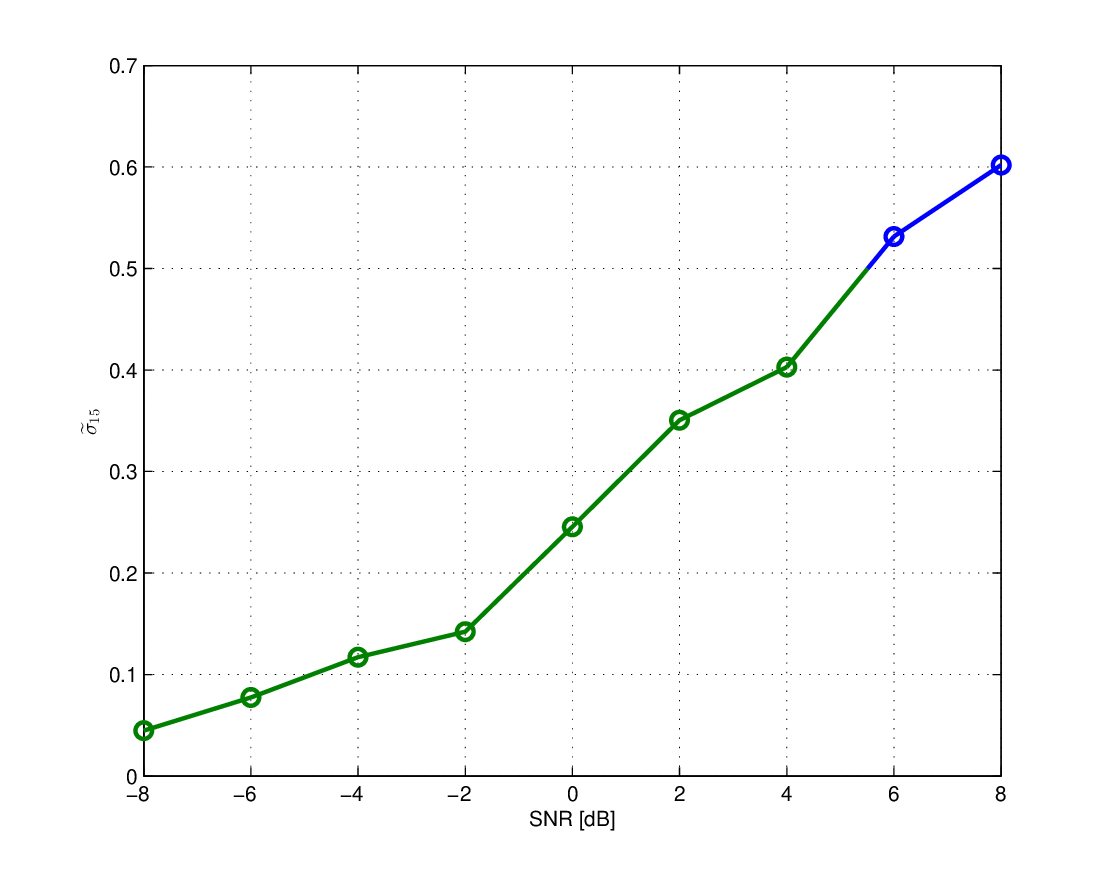}
\caption{Monotonic decrease of $\widetilde{\si}_{15}$ with decreasing $SNR[dB]$ in practical case. The 0.5 threshold is crossed between $SNR[dB]=4$ and $SNR[dB]=6.$}
\label{fig:4}
\end{figure}

Consider now the case where the channel matrix $\H_c$ is assumed to be known at the receiver with an error such that $\widetilde{\H}_c=\H_c+\E_c$, where 
the entries of the error matrix $\E_c$ are i.i.d. drawn from a Gaussian distribution with zero-mean and variance~$10^{-4}.$ It is also assumed below that neither the noise covariance matrix $R_{\n_c}$ nor the noise power~$\ep$ are available at the receiver side, and only  the sample estimate of the covariance matrix of the observed data $\Ry$ is available:
\begin{equation} \label{esteem}
\widehat{\Ry}={{1}\over{Q}}\sum_{q=1}^Q\y_q\y_q^t,
\end{equation}
where $Q=200$ is the number of data blocks. For the results presented in Fig.\ref{fig:3}-\ref{fig:4}, the perturbed channel matrix $\widetilde{\H}_c$ is assumed known, and this matrix is used in place of $\H_c$ which was used for the results presented in Fig.\ref{fig:1}-\ref{fig:2}. Moreover, for both sets of results, the same data block is assumed to be transmitted in order to clearly illustrate the difference between the results obtained under complete and incomplete model knowledge.

\begin{figure}[t]
  \includegraphics[scale=.506]{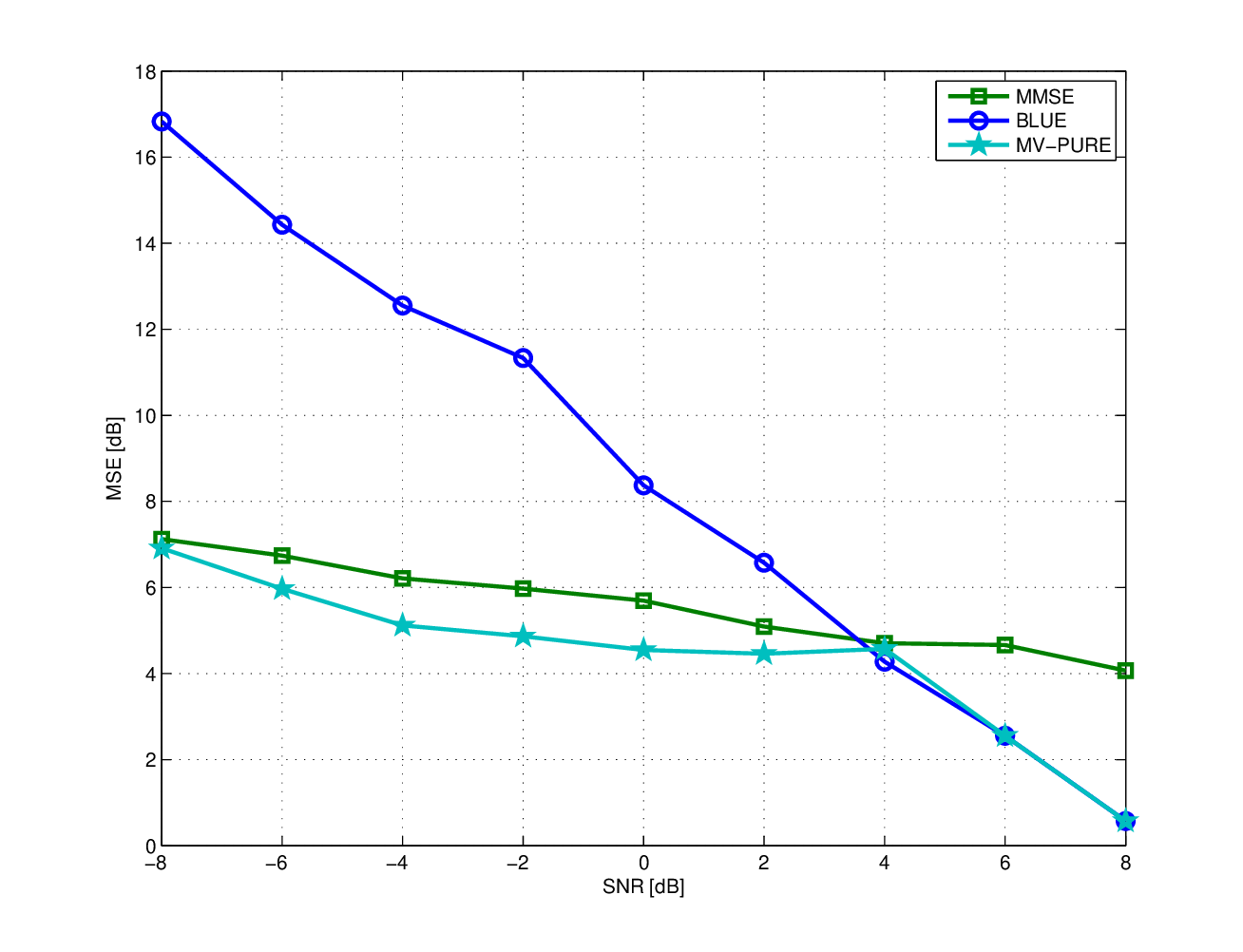}
\caption{MSE [dB] vs. SNR [dB] for a sample channel realization in practical case.}
\label{fig:3}
\end{figure}

In the current settings one cannot use theorem \ref{battery} directly due to unknown $\Rn$ and~$\ep.$ However, the last two trailing eigenvalues $\si_{15},\si_{16}$ of $\H^t\Ry^{-1}\H$ may be estimated via the last two trailing eigenvalues $\widetilde{\si}_{15},\widetilde{\si}_{16}$ of $\widetilde{\H}^t\widehat{\Ry}^{-1}\widetilde{\H}.$ Moreover, similarly as before, one may utilize theorem \ref{blackened} to set $r=14$ for $SNR[dB]=(-8,-6,-4,-2,0,2,4)$ and $r=16$ for $SNR[dB]=(6,8).$ The results are presented in Fig.\ref{fig:3}-\ref{fig:4}, where in Fig.\ref{fig:3} it is seen that the rank choice remains correct in the current settings. This is because the trailing eigenvalues of $\H^t\Ry^{-1}\H$ are not significantly perturbed in $\widetilde{\H}^t\widehat{\Ry}^{-1}\widetilde{\H}$, which is demonstrated in Fig.\ref{fig:4} below. Note that the slight imperfection for $SNR[dB]=4$ in Fig.\ref{fig:3} is due to the single-trial estimate of the MSE.

We note also that the mildly ill-conditioned matrix $\H^t\Rn^{-1}\H$ used in the simulations above implied that the rank-reduction capability of the stochastic MV-PURE estimator provided gain in performance over the stochastic BLUE estimator in the highly noisy settings of $SNR[dB]=(-8,-6,-4,-2,0,2,4)$, \emph{cf.} also the averaged performance (over 10 000 Monte-Carlo runs) demonstrated in \cite{Piotrowski2009}. Indeed, if one considers channel correlation in model (\ref{mul}), which is induced by the propagation environment or spacing between antennas, the resulting channel representation $\H_c$ may be severely ill-conditioned \cite{Gesbert2003, Stoica2003}. This in turn leads to significantly ill-conditioned matrix $\H_c^t\R_{\n_c}^{-1}\H_c$ (and thus also its real-valued representation $\H^t\R_{\n}^{-1}\H$). In such a case, the stochastic MV-PURE estimator may obtain gain in performance over the stochastic BLUE estimator for higher values of SNR due to the interplay between the noise power $\ep$ and ill-conditioning of $\H^t\Rn^{-1}\H$ as discussed below theorem~\ref{battery}.

\section{Concluding remarks} \label{conclusion}
In highly noisy settings, we proved that the stochastic MV-PURE estimator achieves drastic improvement in performance over its full-rank version, the stochastic BLUE estimator. This result demonstrates that many of the existing applications of the stochastic BLUE estimator may benefit by employing instead the reduced-rank approach of the stochastic MV-PURE estimator in highly noisy conditions.

\section*{Acknowledgment} The authors are grateful to anonymous reviewers for their constructive comments which surely promoted the readability of the revised manuscript. They would also like to thank Dr. Renato L. G. Cavalcante for his help on editing this paper.



\bibliographystyle{model1-num-names}
\bibliography{references}







\end{document}